\documentclass[12pt]{article}
\usepackage{a4wide,amssymb,cite}
\pdfoutput=1

\usepackage{a4wide,amssymb,graphicx}
\usepackage{epsfig}
\usepackage[usenames,dvipsnames]{color}
\usepackage{slashed}
\usepackage{amssymb,cite,graphicx}
\usepackage{slashed}
\usepackage{amsmath,bm,bbm}
\usepackage{amsfonts}
\usepackage[titletoc,title]{appendix}
\usepackage[small]{caption}
\usepackage[margin=1in]{geometry}
\usepackage[multiple]{footmisc}
\usepackage{mathtools}
\usepackage{slashed}
\usepackage[nottoc]{tocbibind}
\usepackage{xcolor}

\usepackage{tikz}
\usetikzlibrary{shapes,arrows,positioning,automata,backgrounds,calc,er,patterns}
\usepackage[compat=1.1.0]{tikz-feynman}
\usetikzlibrary{trees}
\usetikzlibrary{decorations.pathmorphing}
\usetikzlibrary{decorations.markings}

\newcommand{\be}{\begin{equation}}
\newcommand{\ee}{\end{equation}}
\newcommand{\bea}{\begin{eqnarray}}
\newcommand{\eea}{\end{eqnarray}}

\def\circa#1{\,\raise.3ex\hbox{$#1$\kern-.75em\lower1ex\hbox{$\sim$}}\,}

\begin{document}

\begin{titlepage}
%
%


%

\begin{centering}
\vspace{1cm}
{\Large {\bf On the Weyl gravity extension of Higgs inflation}} \\

\vspace{1.5cm}

{\bf Shuntaro Aoki$^\sharp$ and Hyun Min Lee$^\dagger$}
\\
\vspace{.5cm}

{\it  Department of Physics, Chung-Ang University, Seoul 06974, Korea.}

\vspace{.5cm}


\end{centering}
\vspace{2cm}

\begin{abstract}
\noindent
We consider the embedding of Higgs inflation with a non-minimal coupling into the Weyl gravity. In this model, the effective current-current interactions from the heavy Weyl gauge field cancel the non-canonical Higgs kinetic term in Einstein frame, so the unitarity problem of the original Higgs inflation becomes less severe. For a simple case where the couplings of the heavy Weyl gauge field appears from the non-minimal couplings to the Ricci curvature scalar in Weyl gravity, we find that the resultant model for Higgs inflation is the same as in the Palatini formulation for Higgs inflation.  The crucial difference of our model from the Palatini formulation for Higgs inflation is that there is a light Weyl gauge field coupled to the Higgs fields.
We also generalize the unitarization of Higgs inflation with general covariant kinetic terms for the dilaton and the Higgs fields, and realize a successful Higgs inflation, interpolating between the Palatini formulation for Higgs inflation and a Higgs-like inflation. 
We also discuss the Higgs mechanism for the light Weyl gauge field with an extra singlet scalar and show some interesting signatures for Higgs physics, such as the overall suppression of Higgs couplings and the direct couplings of the light Weyl gauge field to the Higgs boson.

\end{abstract}

\vspace{3cm}

\begin{flushleft} 
$^\sharp$Email: shuntaro@cau.ac.kr  \\
$^\dagger$Email: hminlee@cau.ac.kr 
\end{flushleft}

\end{titlepage}

\section{Introduction}

Effective field theories have been important tools for particle physics and cosmology, due to the fact that it is enough to keep a finite number of parameters consistent with symmetries at low energies for accurate calculations of observable quantities.  The Standard Model (SM) has been shown to be a consistent effective field theory, which is an efficient and accurate way of describing interactions of fundamental particles, probably up to very high energies, due to the null results for new physics at the Large Hadron Collider (LHC). Nonetheless, there are hints for new physics solving the empirical problems such as neutrino masses, baryon number asymmetry, dark matter, and the theoretical problems such as the hierarchy problem, cosmological constant problem, flavor hierarchy, etc. 

Cosmic inflation requires a slowly rolling extra scalar field, the so called inflaton, in the early era of cosmology, solving the problems for homogeneity, isotropy, relics, etc. Quantum fluctuations of the inflaton set the initial conditions for the standard Big Bang cosmology by generating the seeds for inhomogeneities of the Cosmic Microwave Background (CMB) and the large-scale structures. In usual cases, we need to go beyond the SM with a singlet inflaton and introduce extra couplings of  the inflaton appropriately for reheating process. Although it is a challenging task to find a consistent framework for inflation models, the power spectrum of the observed CMB anisotropies such as Planck data \cite{planck} have put meaningful constraints on the inflation models through the spectral index, the tensor-to-scalar ratio, etc.

The Higgs boson is the only fundamental scalar field in the SM, so it is a good candidate for the inflaton.
Then, it is an interesting possibility to regard the Higgs boson as the inflaton, instead of introducing an extra scalar field beyond the SM.
Higgs inflation \cite{higgsinf} is the minimal extension of the SM with a non-minimal coupling $\xi_H$ of the Higgs fields to gravity, resulting in the consistent predictions for inflation. However, a large non-minimal coupling gives rise to a premature violation of unitarity from the non-canonical Higgs kinetic terms in Einstein frame, so the Higgs inflation is still an effective field theory the validity of which is limited below the unitarity scale, $M_P/\xi_H$ \cite{problem}.   Thus, there are several proposals to unitarize the Higgs inflation in linear sigma models with an extra singlet scalar field  \cite{unitarity,unitarity2,espinosa} or with $R^2$ or higher curvature terms \cite{ema0,gorbunov,ema,general,HiggsR2susy,HiggsR2,R3}.

In this article, we make an attempt to embed the Higgs inflation in the context of Weyl gravity where the conformal symmetry is gauged by a Weyl gauge field \cite{Weylold}.  In this framework, the conformal symmetry is spontaneously broken by the Vacuum Expectation Value (VEV) of the dilaton scalar. As a result, the Planck scale and the Higgs mass parameter are generated dynamically, and the Weyl gauge field coupled to the dilaton becomes heavy \cite{Weyl}. We study the conditions that the effective current-current interactions coming from the Weyl gauge field cancel the non-canonical Higgs kinetic terms in Higgs inflation such that the unitarity scale gets higher than in Higgs inflation. In the cases with one or two Weyl gauge fields, we show how a solution to unitarity problem in Higgs inflation and a successful inflation are achieved at the same time. The couplings of one Weyl gauge field is always absorbed into the non-minimal couplings to the Ricci curvature scalar in Weyl gravity, but those of the other Weyl gauge field is not. We show the interplay of inflationary predictions with the unitarity scale in the case with two Weyl gauge fields.

The paper is organized as follows. 
We first review the unitarity problem in Higgs inflation and identify the non-canonical Higgs kinetic terms in Einstein frame as the current-current interactions for the heavy Weyl gauge field.  Then, we propose the Weyl invariant Lagrangian with non-minimal couplings for the dilaton and the Higgs fields, and discuss the roles of the second Weyl gauge field for unitarization and inflation, in comparison to the case with a single Weyl gauge field. Next we generalize the Weyl invariant Lagrangian with extra kinetic terms for the dilaton and the Higgs fields, which are not absorbed into the the non-minimal couplings to the Ricci curvature scalar in Weyl gravity. We continue to discuss the implications of the light Weyl gauge field for Higgs physics and collider experiments. Finally, conclusions are drawn. 
There are two appendices for the comparison to the unitarization of Higgs inflation with a singlet scalar and the details on the generalized Weyl-invariant Lagrangian.

\section{Higgs inflation and unitarization}

We consider the Higgs inflation with a non-minimal coupling in relation to the unitarity problem and sketch a new way of unitarizing the Higgs inflation in the presence of the heavy Weyl gauge coupled to the Higgs fields.

\subsection{Unitarity problem in Higgs inflation}

The Lagrangian for Higgs inflation \cite{higgsinf} is given by
\bea
{\cal L} =\sqrt{-g} \bigg(-\frac{1}{2} (M^2_P+2\xi_H |H|^2) R +|D_\mu H|^2 - V(H) \bigg)
\eea
where $V(H)$ is the Higgs potential,
\bea
V(H) = m^2_H |H|^2 +\lambda_H |H|^4. 
\eea
Making a Weyl transformation by $g_{E,\mu\nu}= \Omega\, g_{\mu\nu}$ with $\Omega=1+2\xi_H |H|^2/M^2_P$,  we get the Einstein frame Lagrangian as
\bea
{\cal L}_E =\sqrt{-g_E} \bigg(-\frac{M^2_P}{2} R +\frac{1}{\Omega} \,|D_\mu H|^2 +\frac{3\xi^2_H}{M^2_P} \frac{(\partial_\mu |H|^2)^2}{\Omega^2} - \frac{V(H)}{\Omega^2}\bigg).  \label{EinsteinL}
\eea
Then, for $2\xi_H|H|^2\gg M^2_P$, a slow-roll inflation takes place, because the Einstein frame is given by $V_E\simeq \frac{\lambda_H}{4\xi^2_H}M_P^4 (1+M^2_P/(2\xi_H |H|^2))^{-2}$, and the canonical Higgs field identified during inflation is given by $\chi/M_P=\sqrt{\frac{3}{2}}\ln (2\xi_H |H|^2/M_P^2)$, leading to $V_E\simeq \frac{\lambda_H}{4\xi^2_H}M_P^4 \big(1+e^{-\sqrt{\frac{2}{3}} \chi/M_P}\big)^{-2}$. Thus, the inflationary predictions are consistent with Planck data, under the condition that $\xi_H\sim 10^4 \sqrt{\lambda_H}$ is fixed by the CMB normalization. Therefore, we need a large non-minimal coupling $\xi_H\gg 1$ for a sizable $\lambda_H$ during inflation. 

The non-canonical Higgs kinetic term with $\xi_H\gg 1$ in eq.~(\ref{EinsteinL}) contains a dimension-6 operator, $(\partial_\mu |H|^2)^2$, with the cutoff scale given by $\Lambda\sim M_P/\xi_H$, so it leads to a premature violation of unitarity at $\Lambda=M_P/\xi_H$, which is much below the inflation scale, $V_I\sim M_P/\sqrt{\xi_H}$ \cite{problem}. Such a low cutoff scale in the vacuum and during reheating casts doubt on the validity of the semi-classical description of Higgs inflation at large Higgs field values.
For unitarizing the Higgs inflation beyond the unitarity scale, as reviewed in Appendix A,  several extensions of the Higgs inflation, such as linear sigma models \cite{unitarity,unitarity2}, singlet scalars \cite{espinosa}, $R^2$ term \cite{ema0,gorbunov,ema,general,HiggsR2susy,HiggsR2}  and general higher curvature terms \cite{general,R3}, have been proposed.

\subsection{Weyl current interactions for unitarization}

For the non-minimal coupling for the Higgs in the Jordan frame, we take the Noether current for the conformal transformation, given by $K_\mu=\partial_\mu K_H$ with $K_H=12\xi_H |H|^2$. Then,  we can rewrite the non-canonical Higgs kinetic term by the current-current interaction,
\bea
\frac{{\cal L}_{H,{\rm eff}} }{\sqrt{-g_E}}=\frac{3\xi^2_H}{M^2_P} \frac{(\partial_\mu |H|^2)^2}{\Omega^2}= \frac{1}{48M_P^2} \frac{K_\mu K^\mu}{\Omega^2}. \label{Hint}
\eea
The purpose of the following discussion is to propose the necessary Weyl couplings to cancel the above dangerous interactions in Weyl gravity. 

Now we consider a Weyl gauge field with mass $m^2_w=6 g^2_w M^2_P$, with the following Jordan frame Lagrangian,
\bea
\frac{{\cal L}_J}{\sqrt{-g}} = -\frac{1}{4} w_{\mu\nu} w^{\mu\nu} + \frac{1}{2}m^2_w w_\mu w^\mu -\frac{1}{2} g_w w_\mu K^\mu+  \frac{1}{2} g^2_w w_\mu w^\mu K_H. \label{weyl}
\eea
The above form of the Lagrangian can be obtained from a non-minimal coupling of the Higgs to Weyl gravity, as will be discussed in Section 3.1 and later sections. Then, after integrating out the Weyl gauge field with its equation of motion,
\bea
w_\mu = \frac{g_w}{2} \frac{K_\mu}{m^2_w +g^2_w K_H},
\eea
we get the effective interactions for the Higgs as
\bea
\frac{{\cal L}_{J,{\rm eff}}}{\sqrt{-g}} &=& -\frac{g^2_w}{8} \frac{K_\mu K^\mu}{m^2_w +g^2_w K_H} \nonumber \\
&=&-\frac{1}{48 M^2_P} \frac{K_\mu K^\mu}{\Omega}. 
\eea
As a result, the corresponding Einstein frame Lagrangian is
\bea
\frac{{\cal L}_{E,{\rm eff}}}{\sqrt{-g_E}} = -\frac{1}{48 M^2_P} \frac{K_\mu K^\mu}{\Omega^2}. \label{weylint}
\eea
Therefore, the effective interaction coming from the Weyl gauge field in eq.~(\ref{weylint}) cancels exactly the non-canonical Higgs kinetic term in eq.~(\ref{Hint}), so the unitarity problem in Higgs inflation disappears.
In the following, we consider a concrete realization of the effective interactions in Weyl gravity. 
We note that the dimension-6 operator of the form $(\partial_\mu |H|^2)^2$ as in eq.~(\ref{weylint})  was introduced  in Jordan frame for unitarity \cite{counterterm}, but the origin of such a higher dimensional operator was not discussed. In contrast, in our work, we will show for the first time that the counter term in Ref.~\cite{counterterm} is the part of the Weyl invariant Lagrangian by construction.

\section{Minimal Weyl gravity for Higgs inflation}

In this section, we introduce the minimal Weyl invariant Lagrangian with a single Weyl gauge field $w_\mu$ and discuss the Higgs inflation and the unitarity problem in this case.

\subsection{Weyl invariant Lagrangian}

In Weyl gravity, the Weyl gauge field can be introduced as a part of the redefined Christoffel symbols with a Weyl gauge coupling $g_w$ \cite{Weyl,Weyl2} by
 \bea
 {\tilde\Gamma}^\rho_{\mu\nu} = \Gamma^\rho_{\mu\nu} +g_w \Big(\delta^\rho_\mu w_\nu+ \delta^\rho_\nu w_\mu-g_{\mu\nu} w^\rho \Big),
 \eea
 which is Weyl-invariant.
 Thus, the resulting Ricci scalar in Weyl gravity \cite{Weyl,Weyl2} is given in terms of the one in Einstein gravity and the Weyl gauge field contribution from $w_\mu$, as follows,
 \bea
 {\tilde R}({\tilde\Gamma})=R(\Gamma)- 6g_w D_\mu w^\mu -6 g^2_w w^\mu w_\mu, \label{Ricci}
 \eea
 with $D_\mu w^\mu=\partial_\mu w^\mu+\frac{1}{2} w^\rho g^{\lambda\beta} \partial_\rho g_{\lambda\beta}$.
 Consequently, we can construct the minimal Lagrangian with one Weyl gauge field $w_\mu$ and the Higgs, as follows,
  \bea
\frac{{\cal L}_{\rm min}}{\sqrt{-g}} = -\frac{1}{2}  (\xi_\phi \phi^2+2 \xi_H |H|^2) {\tilde R}({\tilde\Gamma})-\frac{1}{4} w_{\mu\nu} w^{\mu\nu} - V(H,\phi), \label{minimala}
 \eea
 which is Weyl invariant under the following transformations,
 \bea
 g_{\mu\nu} \to  e^{2\alpha} g_{\mu\nu}, \qquad \phi\to e^{-\alpha}\phi, \qquad H \to e^{-\alpha} H,\qquad w_\mu \to w_\mu -\frac{1}{g_w}\, \partial_\mu \alpha.
 \eea
 Here, the scalar potential is given by
 \bea
 V(H,\phi) =\frac{1}{2} \lambda_{\phi H}\phi^2 |H|^2 + \frac{1}{4}\lambda_\phi \phi^4 +\lambda_H |H|^4.
 \eea

 The Lagrangian in eq.~(\ref{minimala}) is expanded, as follows,
 \bea
 \frac{{\cal L}_{\rm min}}{\sqrt{-g}} &=& -\frac{1}{2} \xi_\phi \bigg(\phi^2 R +6(\partial_\mu\phi)^2 - 6(D_\mu \phi)^2 \bigg) -\frac{1}{4} w_{\mu\nu} w^{\mu\nu}  \nonumber \\
&&-\xi_H \bigg(|H|^2 R+ 6 |\partial_\mu H|^2-6 |D_\mu H|^2 \bigg)- V(H,\phi).  \label{minimal1}
\eea
Here, we rewrote the Weyl current interactions to the dilaton and Higgs fields through the non-minimal couplings by integration by parts, and absorbed them in the Weyl covariant derivative terms, $(D_\mu\phi)^2$ and $|D_\mu H|^2$.
However, in this case, there is no net kinetic term for Higgs or dilaton in the Jordan frame. As a result, $\phi^2 R +6(\partial_\mu\phi)^2 $ and $|H|^2 R+ 6 |\partial_\mu H|^2$ are Weyl invariant, respectively, so the Weyl invariance is manifest in Einstein gravity.

Therefore, we need to extend the minimal case in eq.~(\ref{minimal1}) by introducing the Higgs kinetic term in the Jordan frame in the following way. Using eq.~(\ref{weylinvg}) with a single Weyl gauge field $w_\mu$ in Appendix B, we can generalize the above Lagrangian in eq.~(\ref{minimal1}) to the general Weyl invariant Lagrangian \cite{Weyl}, 
\bea
\frac{{\cal L}_1}{\sqrt{-g}} &=& -\frac{1}{2} \xi_\phi \bigg(\phi^2 R +6(\partial_\mu\phi)^2 - 6r_\phi (D_\mu \phi)^2 \bigg) -\frac{1}{4} w_{\mu\nu} w^{\mu\nu}  \nonumber \\
&&-\xi_H \bigg(|H|^2 R+ 6 |\partial_\mu H|^2-6 r_H  |D_\mu H|^2 \bigg)  - V(H,\phi). \label{minimal2}
\eea
Here, we note that $r_\phi$ and $r_H$ parameterize the relative strength of the gravitational couplings for two Weyl invariants, $\phi_i^2 R+6(\partial_\mu \phi_i)^2$ and $(D_\mu\phi_i)^2$, so they cannot be eliminated by the field redefinitions. But, we need to take $r_H=1+\frac{1}{6\xi_H}$ for the canonical kinetic term for the Higgs fields in Jordan frame in the following discussion.

We remark that Weyl symmetry appears anomalous at the quantum level in usual dimensional regularization or cutoff regularization where an explicit mass scale is introduced. However, Weyl symmetry can be respected at the quantum level in the presence of quantum scale invariance \cite{quantumscale,Weyl}. In this case, the renormalization scale $\mu$ is replaced by the dilaton scalar $\phi$ at the expense of non-renormalizable dilaton interactions and the loop corrections can be included systematically while preserving the Weyl symmetry.  

Fixing the gauge by $\langle\phi^2\rangle=M^2_P/\xi_\phi$, from eq.~(\ref{minimal2}), we obtain the gauge-fixed Lagrangian as follows,
\bea
\frac{{\cal L}_2}{\sqrt{-g}} &=& -\frac{1}{2}  (M^2_P+2\xi_H |H|^2) R + |\partial_\mu H|^2 -V(H) \nonumber \\
&&-\frac{1}{4} w_{\mu\nu} w^{\mu\nu} + \frac{1}{2}m^2_w w_\mu w^\mu -\frac{1}{2} g_w w_\mu K^\mu+  \frac{1}{2} g^2_w w_\mu w^\mu K_H
\eea
with
\bea
V(H) =\frac{\lambda_\phi M^4_P}{4\xi^2_\phi}+\frac{\lambda_{\phi H} M^2_P}{2\xi_\phi} \, |H|^2 + \lambda_H |H|^2.
\eea
The second line in the above Lagrangian takes the same form as in eq.~(\ref{weyl}), except that $m^2_w=6 r_\phi  g^2_w M^2_P$ and $K_H$ is now modified to $K_H=12r_H\xi_H|H|^2= 2(1+6\xi_H)|H|^2$, instead of $K_H=12\xi_H |H|^2$. 
Here, we can choose a very small $\lambda_\phi$ for the observed cosmological constant and the effective Higgs mass is determined  by $m^2_H=\frac{\lambda_{\phi H}}{2\xi_\phi}\, M_P^2$. For electroweak symmetry breaking, we need to choose $\lambda_{\phi H}<0$ and $|\lambda_{\phi H}/(2\xi_\phi)|\ll 1$.

\subsection{Einstein frame Lagrangian}

Taking the same procedure as in Appendix B (namely, the conformal transformation to Einstein frame and the field redefinition of the Weyl gauge field), we obtain the  Einstein frame Lagrangian after a gauge fixing, $\langle\phi^2\rangle=M^2_P/\xi_\phi$, as follows,
\bea
\frac{{\cal L}_E}{\sqrt{-g_E}} &=&-\frac{M^2_P}{2} R+6\xi_H(r_H-1) \frac{|\partial_\mu H|^2}{\Omega} +\frac{3\xi^2_H}{M^2_P} \frac{(\partial_\mu |H|^2)^2}{\Omega^2}- \frac{V(H)}{\Omega^2} \nonumber \\
&&-\frac{1}{4} {\tilde w}_{\mu\nu} {\tilde w}^{\mu\nu}+ \frac{1}{2}(m^2_w + g^2_w r_H K_H)  \Omega^{-1} {\tilde w}_\mu {\tilde w}^\mu-  \frac{ g^2_w r^2_H}{8\Omega} \,\frac{K_\mu K^\mu}{m^2_w + g^2_w r_H K_H}
\eea
where ${\tilde w}_\mu$ is similarly redefined as in eq.~(\ref{wredef}), but with the modified $K_H$.
This result gives rise to the Lagrangian relevant for inflation, as follows, 
\bea
\frac{{\cal L}_{E,{\rm inf}}}{\sqrt{-g_E}}&=&-\frac{M_P^2}{2} R+\frac{1}{\Omega}(6\xi_H(r_H-1)) |\partial_\mu H|^2 \nonumber \\
&&+\frac{3\xi^2_H}{M^2_P \Omega^2} \frac{r_\phi-r^2_H+2r_H(1-r_H)\xi_H|H|^2/M^2_P}{r_\phi+2r_H \xi_H|H|^2/M^2_P}\, (\partial_\mu|H|^2)^2 -\frac{V(H)}{\Omega^2}   \nonumber \\
&=&-\frac{M_P^2}{2} R+\frac{1}{\Omega} |\partial_\mu H|^2\nonumber \\
&& +\frac{1}{M^2_P \Omega^2} \frac{3r_\phi\xi^2_H-3(\xi_H+\frac{1}{6})^2-\xi_H(\xi_H+\frac{1}{6})|H|^2/M^2_P}{r_\phi+2(\xi_H+\frac{1}{6})|H|^2/M^2_P}\, (\partial_\mu|H|^2)^2
- \frac{V(H)}{\Omega^2} \label{WeylLnew}
\eea
where we chose $r_H=1+\frac{1}{6\xi_H}$ for the Higgs fields in Jordan frame to take a canonical form.
Thus, expanding the above Lagrangian for the Higgs about zero, we obtain the leading higher dimensional terms,
\bea
\frac{{\cal L}_{E,{\rm inf}}}{\sqrt{-g_E}} \supset& &-\frac{2\xi_H}{M^2_P} |H|^2 |\partial_\mu H|^2+\frac{3}{M^2_P} \Big(\xi^2_H-\frac{1}{r_\phi}\Big(\xi_H+\frac{1}{6}\Big)^2 \Big) (\partial_\mu |H|^2)^2 \nonumber \\
&&=\frac{3}{M^2_P} \Big(\xi^2_H -\frac{1}{r_\phi}\Big(\xi_H+\frac{1}{6}\Big)^2+\frac{\xi_H}{3}\Big) (\partial_\mu |H|^2)^2+\cdots \label{Ekinetic}
\eea
where integration by parts are made, and the equation of motion for the Higgs fields, namely, $\Box H\simeq-\frac{\partial V_E}{\partial H^\dagger}$, with $V_E=V(H)/\Omega^2$, is used in the second line. Here, we note that there also appear extra higher dimensional terms from the potential, but they are suppressed by powers of $M_P/\sqrt{\xi_H}$.
Therefore, from eq.~(\ref{Ekinetic}), the unitarity scale is identified from the leading dimension-6 operator by
\bea
\Lambda_1=\frac{M_P}{ \Big|\xi_H(3\xi_H+1)\big(1-\frac{1}{r_\phi}\big)-\frac{1}{12r_\phi}\Big|^{1/2}},
\eea
which becomes of order the Planck scale for $r_\phi=1$, independent of $\xi_H$. Indeed, for $r_\phi=1$, the Ricci curvature scalar for the Higgs field metric becomes constant, $R(H)=-2/M_P^2$, so the Higgs fields live on a four-dimensional hyperbolic space with the curvature of order the Planck scale. 
However, the higher dimensional terms from the scalar potential lead to the cutoff scale, $\Lambda=M_P/\sqrt{\xi_H}$, which is still much larger than the one in Higgs inflation.

\subsection{Inflation}

We consider the case with $r_\phi=r_H=1+\frac{1}{6\xi_H}$ for simplicity.
Then,  the Weyl gauge field is decoupled from the Higgs fields, and the Lagrangian in eq.~(\ref{WeylLnew}) gets simplified to
\bea
\frac{{\cal L}_{E,{\rm inf}}}{\sqrt{-g_E}} =-\frac{M_P^2}{2} R+\frac{1}{\Omega} |\partial_\mu H|^2- \frac{ \xi_H}{2M^2_P\Omega^2}(\partial_\mu|H|^2)^2 - \frac{V(H)}{\Omega^2}. \label{minimal}
\eea
We note that the unitarity scale is given by $\Lambda=M_P/\sqrt{\xi_H}$ for $\xi_H\gtrsim 1$.
In unitary gauge for the Higgs fields, eq.~(\ref{minimal}) becomes
\bea
\frac{{\cal L}_{E,{\rm inf}}}{\sqrt{-g_E}} =-\frac{M_P^2}{2} R+\frac{ (\partial_\mu h)^2}{2(1+\xi_H h^2/M^2_P)^2}- \frac{V(h)}{(1+\xi_H h^2/M^2_P)^2}. \label{minimal0}
\eea
Taking $h\gg M_P/\sqrt{\xi_H}$ during inflation, we find that the coefficient of the Higgs kinetic term is dominated by $ 1/h^4$, so the canonical field $\chi$ is approximated to $\chi\simeq -M^2_P/(\xi_Hh)$. Thus, the inflaton potential in Einstein frame becomes 
\bea
V_E= \frac{\lambda_H M_P^4}{4\xi^2_H} \Big(1+\frac{M^2_P}{\xi_H h^2}\Big)^{-2}\simeq \frac{\lambda_H M_P^4}{4\xi^2_H}\Big(1+\frac{\xi_H \chi^2}{M^2_P}\Big)^{-2}.
\eea 
In this case, we need to choose a very small $\xi_H$ for a successful inflation \cite{Weyl}, so it belongs to a different class of inflation models where there is no unitarity problem below the Planck scale from the beginning. Similarly, for $r_\phi\neq r_H$, the Weyl gauge field is not decoupled from the Higgs. However, as far as the Weyl gauge field is sufficiently heavier than the Hubble scale during inflation,  a slow-roll inflation can be still realized for a very small $\xi_H$ even for $r_\phi\neq r_H$ as in Ref.~\cite{Weyl}. 

We conclude this section by saying that the minimal Weyl gravity with a very small $\xi_H$ is an interesting possibility without a unitarity problem below the Planck scale. However, we would need a very small $\lambda_H$ to get the correct CMB normalization, which requires a severe fine-tuning of the low-energy parameters for a small running Higgs quartic coupling at the inflation scale. Moreover, in our work, we are interested in the dynamical mechanism for unitarizing the original Higgs inflation with a large non-minimal coupling. So, we consider a possibility of extending the Weyl symmetry in the next sections.

\section{Higgs inflation with extended Weyl symmetry}

We consider the Weyl invariant Lagrangian with the dilaton and the Higgs fields, in the presence of non-minimal couplings for them. We introduce the non-minimal couplings to the modified Ricci curvature scalar in Weyl gravity which contains the couplings of the Weyl gauge fields, and add an extra kinetic term for the Higgs fields with an extra Weyl gauge field. 
In this case, we discuss the unitarity scale and the inflationary predictions.

 \subsection{Lagrangian with two Weyl gauge fields}

For a concrete realization of unitarization and successful Higgs inflation with a large non-minimal coupling, we consider the Lagrangian including the dilaton $\phi$ and two Weyl gauge fields, $w_\mu$ and $X_\mu$, in the Jordan frame. Another Weyl gauge field $X_\mu$ does not appear in the Ricci scalar but it gives rise to an extra Weyl-invariant kinetic term for the Higgs field.

We first discuss the origin of two Weyl gauge fields.
Since the Weyl gauge fields are originated from the local scale symmetries, we need to consider the doubled coordinates, $x^\mu_1$ and $x^\mu_2$, leading to the extended diffeomorphism invariance in Weyl gravity, ${\rm Diff}_1\times {\rm Diff}_2$, with two metric tensors, $g_{1,\mu\nu}$ and $g_{2,\mu\nu}$, and the corresponding Christoffel symbols, ${\tilde\Gamma}^\rho_{1,\mu\nu}$ and ${\tilde\Gamma}^{\rho}_{2,\mu\nu}$, Weyl gauge fields, $w_{1,\mu}$ and $w_{2,\mu}$, as well as the dilaton scalars, $\phi_1$ and $\phi_2$. Then, including a pair of Higgs fields, $H_1$ and $H_2$, with the non-minimal couplings in the extended Weyl gravity, we can generalize the action for the conformal sector in eq.~(\ref{minimala})  to 
\bea
S&=&\sum_{i=1,2} \int dv_i \bigg[ -\frac{1}{2} (\xi_i \phi^2_i+2\zeta_i|H_i|^2) {\tilde R}({\tilde\Gamma}_i) -\frac{1}{4} w_{i,\mu\nu} w^{\mu\nu}_i \bigg]+\Delta S
\eea
where $dv_i\equiv d^4x_i  \sqrt{-g_i}$ are the infinitesimal spacetime volumes, and $\xi_i, \zeta_i (i=1,2)$ are the non-minimal couplings for the dilaton and Higgs fields, and the Ricci scalar in Weyl gravity is decomposed into ${\tilde R}({\tilde\Gamma}_i)=R(\Gamma_i)- 6g_{w_i} D_\mu w^\mu_i -6 g^2_{w_i} w^\mu_i w_{i,\mu}$ with $g_{w_i}(i=1,2)$ being the Weyl gauge couplings, and $\Delta S$ are the interaction terms between the Weyl gauge fields and the general covariant derivative terms for the Higgs fields, given by
\bea
\Delta S &=&\sum_{i=1,2} \int dv_i \, (-3a_i \xi_i \phi^2_i) \Big(g_{w_1} w_{1,\mu} +\kappa_i g_{w_2} w_{2,\mu}\Big)\Big(g_{w_1} w_{1,\nu} +\kappa_i g_{w_2} w_{2,\nu}\Big)g^{\mu\nu}_i  \nonumber \\
& +&\sum_{i=1,2} \int dv_i \, (-6{\hat a}_i\zeta_i |H_i|^2) \Big(g_{w_1} w_{1,\mu} +{\hat\kappa}_i g_{w_2} w_{2,\mu}\Big)\Big(g_{w_1} w_{1,\nu} +{\hat\kappa}_i g_{w_2} w_{2,\nu}\Big) g^{\mu\nu}_i \nonumber \\
&+&\sum_{i=1,2} \int dv_i \Big(\partial_\mu- b_i g_{w_1} w_{1,\mu}- c_i g_{w_2} w_{2,\mu} \Big)H^\dagger_i\Big(\partial_\nu- b_i g_{w_1} w_{1,\nu}- c_i g_{w_2} w_{2,\nu} \Big)H_i g^{\mu\nu}_i
\label{Wmixing}
\eea
with $a_i,  {\hat a}_i, \kappa_i, {\hat\kappa}_i, b_i,  c_i (i=1,2)$ being extra couplings.  
Then, the action $S$ is Weyl-invariant under
\bea
g_{i,\mu\nu} \to  e^{2\alpha_i(x_i)} g_{i,\mu\nu}, \quad \phi_i\to e^{-\alpha_i(x_i)}\phi_i,  \quad  H_i\to e^{-\alpha_i(x_i)} H_i,  \quad w_{i,\mu} \to w_{i,\mu} -\frac{1}{g_{w_i}}\,\frac{\partial\alpha_i(x_i)}{\partial x^\mu_i},
\eea  
with $\alpha_i(i=1,2)$ being two independent Weyl transformation parameters.
On the other hand, the extra action $\Delta S$ with generic parameters breaks the extended diffeomorphism invariance as well as the Weyl invariance explicitly. However, after the doubled coordinates are identified by $x^\mu_1=x^\mu_2$, $\Delta S$ becomes diffeomorphism invariant, as well as Weyl-invariant, provided that 
\bea
\alpha_1+\kappa_i \alpha_2=0, \qquad \alpha_1+{\hat\kappa}_i \alpha_2=0,  \qquad i=1,2,
\eea 
and 
\bea
b_i \alpha_1 + c_i  \alpha_2 =\alpha_i, \qquad i=1,2.
\eea

Suppose that the full diffeomorphism invariance is broken by ${\rm Diff}_1\times {\rm Diff}_2\to {\rm Diff}_0$, such that two coordinates, two metric tensors and the dilaton scalars are identified by $x^\mu_1=x^\mu_2$, $g_{1,\mu\nu}=g_{2,\mu\nu}\equiv g_{\mu\nu}$ and $\phi_1=\phi_2\equiv \phi/\sqrt{2}$, respectively.  Moreover, we also identify two Higgs fields by $H_1=H_2\equiv H/\sqrt{2}$.
Furthermore, taking $\xi_1=\xi_2\equiv \xi_\phi$ and $\zeta_1=\zeta_2\equiv \xi_H$, and imposing the following conditions for the parameters in  the extra action $\Delta S$ in eq.~(\ref{Wmixing}),
\bea
a_1+a_2 =a_1\kappa^2_1 + a_2 \kappa^2_2=-a_1 \kappa_1 - a_2 \kappa_2 = \frac{1}{2}, \label{conditions}
\eea
with ${\hat a}_i=a_i$ and ${\hat\kappa}_i=\kappa_i$ ($i=1,2$), 
and
\bea
b_1=b_2=\frac{g_{w_2}g_X}{2 g_{w_1}g_w}, \quad  c_1=c_2=-\frac{g_{w_1}g_X}{2 g_{w_2}g_w}, \label{conditions2}
\eea
we obtain the effective Weyl gravity Lagrangian in terms of two redefined Weyl gauge fields, $w_\mu$ and $X_\mu$, as follows,
\bea
{\cal L}_{\rm eff}=\sqrt{-g} \bigg[-\frac{1}{2} (\xi_\phi \phi^2+2\xi_H |H|^2) {\tilde R}({\tilde\Gamma}) -\frac{1}{4}w_{\mu\nu} w^{\mu\nu} -\frac{1}{4} X_{\mu\nu}X^{\mu\nu}+|D'_\mu H|^2  \bigg] \label{effW}
\eea
where the Ricci scalar in the effective Weyl gravity is given  by  ${\tilde R}({\tilde\Gamma})=R(\Gamma)- 6g_w D_\mu w^\mu -6 g^2_{w} w^\mu w_{\mu}$, in terms of the redefined Christoffel symbols,
\bea
{\tilde \Gamma}^\rho_{\mu\nu} &=& \frac{1}{2}({\tilde \Gamma}^\rho_{1,\mu\nu}+{\tilde \Gamma}^\rho_{2,\mu\nu}) \nonumber \\
&=&\Gamma^\rho_{\mu\nu} +g_w \Big(\delta^\rho_\mu w_\nu+ \delta^\rho_\nu w_\mu-g_{\mu\nu} w^\rho \Big),
\eea
Here, $g_w=\frac{1}{2} \sqrt{g^2_{w_1}+g^2_{w_2}}$, and the redefined Weyl gauge fields are given by $w_\mu=(g_{w_1} w_{1,\mu}+g_{w_2} w_{2,\mu})/\sqrt{g^2_{w_1}+g^2_{w_2}}$ and $X_\mu=(g_{w_2} w_{1,\mu}-g_{w_1} w_{2,\mu})/\sqrt{g^2_{w_1}+g^2_{w_2}}$.
We note that there are three equations for four unknown parameters, $a_i, \kappa_i(i=1,2)$, in eq.~(\ref{conditions}), so generically there is one parameter family of the solutions realizing the effective Weyl gravity. Moreover, the covariant derivative with  $X_\mu$ for the Higgs fields is given $D'_\mu H=(\partial_\mu -g_X X_\mu) H$ with an independent gauge coupling $g_X$. 

We remark that the Weyl transformations of the redefined Weyl gauge fields, $w_\mu$ and $X_\mu$, become
\bea
w_\mu &\to& w_\mu -\frac{1}{\sqrt{g^2_{w_1}+g^2_{w_2}}}\, \partial_\mu(\alpha_1+\alpha_2), \\
X_\mu &\to& X_\mu -\frac{1}{\sqrt{g^2_{w_1}+g^2_{w_2}}}\, \partial_\mu \bigg(\frac{g_{w_2}}{g_{w_1}}\, \alpha_1-\frac{g_{w_1}}{g_{w_2}}\, \alpha_2 \bigg).
\eea
As a result, we can identify the remaining Weyl symmetry in the effective Lagrangian in eq.~(\ref{effW}) by
\bea
 w_\mu \to w_\mu -\frac{1}{g_w}\,\partial_\mu \alpha, \qquad X_\mu\to X_\mu -\frac{1}{g_X} \partial_\mu\alpha, \label{Weyl0}
\eea
 with the Weyl transformation parameter being identified as
\bea
\alpha=\frac{1}{2} (\alpha_1+\alpha_2)
\eea
and
\bea
\alpha_2= \frac{g_{w_2}/g_{w_1}-g_X/g_w}{g_{w_1}/g_{w_2}+g_X/g_w}\, \alpha_1.
\eea
Thus, both of two Weyl transformations are nontrivial, as far as $|g_X/g_w|\neq |g_{w_2}/g_{w_1}|$.

We remark on the extra Weyl-invariant terms with two Weyl gauge fields, $w_\mu$ and $X_\mu$, in the effective Weyl gravity.
First, Weyl gauge self-interactions, $(g_w w_\mu-g_X X_\mu)^n$ with $n=2$ and $n\geq 6$, are Weyl-gauge invariant under the Weyl transformations in eq.~(\ref{Weyl0}), but they violate the scale symmetry of the gravitational Lagrangian. On the other hand, there is a Weyl invariant self-interaction, $(g_w w_\mu-g_X X_\mu)^4$, which could lead to new interactions between $X_\mu$ and the SM Higgs after the heavy Weyl gauge field $w_\mu$ is integrated out. But, they are suppressed by the mass of the heavy Weyl photon mass, so  we don't include it in the following discussion. 
Furthermore, we could construct the Wely-invariant terms with dilaton and Higgs fields by $\phi^2(g_w w_\mu-g_X X_\mu)^2$ and $|H|^2(g_w w_\mu-g_X X_\mu)^2$.
However, those terms are absent by construction in the extended Weyl gravity, for appropriate choices of the parameters in the original Lagrangian in eq.~(\ref{Wmixing}) with eqs.~(\ref{conditions}) and (\ref{conditions2}). 
Thus, it is sufficient to take the extra Higgs kinetic term respecting the remnant of the Weyl symmetry under $X_\mu$, as introduced in eq.~(\ref{effW}).
More general cases with covariant derivative terms for the dilaton and the Higgs fields will be discussed in the next section.

Consequently, for the later discussion, we consider a Weyl invariant Lagrangian with two Weyl gauge fields, $w_\mu$ and $X_\mu$, in the effective Weyl gravity with a single metric tensor, in the following simple form, 
 \bea
\frac{{\cal L}_2}{\sqrt{-g}} &=& -\frac{1}{2}  (\xi_\phi \phi^2+2 \xi_H |H|^2) {\tilde R}({\tilde\Gamma})-\frac{1}{4} w_{\mu\nu} w^{\mu\nu}  \nonumber \\
 &&- \frac{1}{4} X_{\mu\nu} X^{\mu\nu}  + |D'_\mu H|^2 - V(H,\phi). \label{weylinv2}
 \eea
The above Lagrangian is invariant under the local conformal transformation with
 \bea
 g_{\mu\nu} \to  e^{2\alpha} g_{\mu\nu}, \qquad \phi\to e^{-\alpha}\phi, \qquad H \to e^{-\alpha} H,
 \eea
 and eq.~(\ref{Weyl0}).
We note that both Weyl gauge fields, $w_\mu$ and $X_\mu$, transform with only one gauge transformation parameter $\alpha$. Since the Weyl gauge field $X_\mu$ couples to gravity minimally,  it can survive at low energy and it receives mass from the VEVs of Higgs and extra light singlet scalars, as will be discussed in Section 5.

After expanding the Weyl-invariant Lagrangian in eq.~(\ref{weylinv2}), we get 
\bea
\frac{{\cal L}_2}{\sqrt{-g}} &=& -\frac{1}{2} \xi_\phi \bigg(\phi^2 R +6(\partial_\mu\phi)^2 - 6(D_\mu \phi)^2 \bigg) -\frac{1}{4} w_{\mu\nu} w^{\mu\nu}  \nonumber \\
&&-\xi_H \bigg(|H|^2 R+ 6 |\partial_\mu H|^2-6 |D_\mu H|^2 \bigg) \nonumber \\
 &&- \frac{1}{4} X_{\mu\nu} X^{\mu\nu}  + |D'_\mu H|^2 - V(H,\phi) \label{weylinv}
\eea
where the Weyl covariant derivatives are defined as $D_\mu\phi=(\partial_\mu -g_w w_\mu)\phi$, $D_\mu H=(\partial_\mu -g_w w_\mu)H$.
Fixing the gauge to $\langle\phi^2\rangle=M^2_P/\xi_\phi$. Then, from eq.~(\ref{weylinv}), we obtain the gauge-fixed Lagrangian as follows,
\bea
\frac{{\cal L}_2}{\sqrt{-g}} &=& -\frac{1}{2}  (M^2_P+2\xi_H |H|^2) R + |D'_\mu H|^2 -V(H)- \frac{1}{4} X_{\mu\nu} X^{\mu\nu}   \nonumber \\
&&-\frac{1}{4} w_{\mu\nu} w^{\mu\nu} + \frac{1}{2}m^2_w w_\mu w^\mu -\frac{1}{2} g_w w_\mu K^\mu+  \frac{1}{2} g^2_w w_\mu w^\mu K_H.
\eea
Thus, in this case, we realize the exactly same Jordan-frame Lagrangian for the Weyl gauge field as in eq.~(\ref{weyl}), with $m^2_w=6 g^2_w M^2_P$  and $K_H=12\xi_H |H|^2$. 

Making a field redefinition for the Weyl gauge field by
\bea
{\tilde w}_\mu = w_\mu-\frac{1}{2g_w}\partial_\mu \ln (m^2_w+ g^2_w K_H),
\eea
we can rewrite the gauge-fixed Lagrangian as
\bea
\frac{{\cal L}_2}{\sqrt{-g}} &=& -\frac{1}{2}  (M^2_P+2\xi_H |H|^2) R +  |D'_\mu H|^2 -V(H)- \frac{1}{4} X_{\mu\nu} X^{\mu\nu}   \nonumber \\
&&-\frac{1}{4} {\tilde w}_{\mu\nu} {\tilde w}^{\mu\nu} + \frac{1}{2}(m^2_w + g^2_w K_H){\tilde w}_\mu {\tilde w}^\mu -  \frac{ g^2_w}{8} \,\frac{K_\mu K^\mu}{m^2_w + g^2_w K_H}.  \label{gf2}
\eea

\subsection{Einstein-frame Lagrangian and unitarity}

Making a Weyl transformation of the metric, $g_{E,\mu\nu}=\Omega g_{\mu\nu}$, with the frame function,
\bea
\Omega=1+2\xi_H |H|^2/M^2_P,
\eea 
and using $m^2_w+ g^2_w K_H=6g^2_w M^2_P \Omega$ for $m^2_w=6 g^2_w M^2_P$, we get the Einstein-frame  Lagrangian from eq.~(\ref{gf2}) as
\bea
\frac{{\cal L}_E}{\sqrt{-g_E}} &=&-\frac{M^2_P}{2} R + \frac{ |D'_\mu H|^2}{\Omega} +\frac{3\xi^2_H}{M^2_P} \frac{(\partial_\mu |H|^2)^2}{\Omega^2} - \frac{V(H)}{\Omega^2}- \frac{1}{4} X_{\mu\nu} X^{\mu\nu}  \nonumber \\
&&-\frac{1}{4} {\tilde w}_{\mu\nu} {\tilde w}^{\mu\nu} + \frac{1}{2}m^2_w {\tilde w}_\mu {\tilde w}^\mu- \frac{1}{48M^2_P}\frac{K_\mu K^\mu}{\Omega^2}.
\eea
Consequently, from $K_\mu=12 \xi_H \partial_\mu |H|^2$, the above Einstein-frame Lagrangian becomes
\bea
\frac{{\cal L}_E}{\sqrt{-g_E}} =-\frac{M^2_P}{2} R+ \frac{ |D'_\mu H|^2 }{\Omega} - \frac{V(H)}{\Omega^2}- \frac{1}{4} X_{\mu\nu} X^{\mu\nu}-\frac{1}{4} {\tilde w}_{\mu\nu} {\tilde w}^{\mu\nu} + \frac{1}{2}m^2_w {\tilde w}_\mu {\tilde w}^\mu. \label{Einsteinsimple}
\eea
As a result, the non-canonical kinetic terms for the Higgs field containing $\xi^2_H$ are cancelled out, taking the same form as in the Palatini formulation for Higgs inflation \cite{Bauer:2008zj,Takahashi:2018brt,Jinno:2019und}.
Moreover, the redefined Weyl gauge field ${\tilde w}_\mu$ is decoupled from the Higgs field.
Therefore, from the remaining Higgs kinetic term and the scalar potential, we can identify the unitarity cutoff as $\Lambda=M_P/\sqrt{\xi_H}$ in the vacuum, which is much larger than the one in the original Higgs inflation, $\Lambda=M_P/\xi_H$, for $\xi_H\gg 1$.

We remark that the crucial difference from the Palatini formulation is that there exists a light Weyl gauge field $X_\mu$, which gets mass only from the VEVs of the Higgs and extra singlet scalars and has interesting implications for Higgs physics, as will be discussed in the later section.

\subsection{Inflation}

Taking the unitary gauge for the Higgs fields by $H=\frac{1}{\sqrt{2}}(0,h)^T$ in eq.~(\ref{Einsteinsimple}),  the part of the Einstein-frame Lagrangian relevant for inflation is given by
\bea
\frac{{\cal L}_{E,{\rm inf}}}{\sqrt{-g_E}} =-\frac{M_P^2}{2} R+ \frac{1}{2}\frac{(\partial_\mu h)^2}{(1+\xi_H h^2/M_P^2)}   - \frac{V(h)}{(1+\xi_H h^2/M_P^2)^2}.
\eea

We assume that the Higgs quartic term is dominant during inflation and $ \xi_H>0$. 
Then, we first make a field redefinition for the canonical field $\chi$ by
\begin{align}
\frac{dh}{\sqrt{1+\xi_Hh^2/M_P^2}} = d \chi,  
\end{align}
which gives rise to
\begin{align}
\chi/M_P=\frac{1}{\sqrt{\xi_H}}\ln \left[\sqrt{\xi_H} h/M_P+\sqrt{1+\xi_Hh^2/M_P^2}\right],
\end{align}
or
\begin{align}
h/M_P=\frac{1}{\sqrt{\xi_H}}{\rm{sinh}}\left[\sqrt{\xi_H}\chi/M_P\right].
\end{align}
Then, the Einstein frame potential becomes
\begin{align}
V_E= \frac{\lambda_H}{4\xi_{H}^{2}} M_P^4\tanh ^{4}\left[\sqrt{\xi_H} \chi/M_P\right].  \label{inf_potential}
\end{align}

The slow-roll parameters, $\epsilon$ and $\eta$, and the number of efoldings $N$, are defined as
\begin{align}
&\epsilon=\frac{M_P^2}{2}\left(\frac{d V_E / d \chi}{V_E}\right)^{2}, \\
&\eta=M_P^2\frac{d^{2} V_E / d \chi^{2}}{V_E},\\
&N=\frac{1}{M_P}\int_{\chi_e}^{\chi} \frac{d \chi}{\sqrt{2 \epsilon}}
\end{align} 
where $\chi_e$ is the field value at the end of inflation, and the CMB normalization, the spectral index and the tensor-to-scalar ratio are respectively given by 
\begin{align}
A_{s}=\frac{1}{24 \pi^{2}} \frac{V_E/M_P^4}{\epsilon}, \quad n_{s}=1-6 \epsilon+2 \eta, \quad r=16 \epsilon.   \label{obs}
\end{align}
Then, we take $\xi_H h^2\gg M_P^2$ for which we get $h/M_P\simeq \frac{1}{\sqrt{\xi_H}}\, e^{\sqrt{\xi_H}\chi/M_P}$ and the inflationary parameters become
\bea
N &\simeq& \frac{1}{32 \xi_{H}} e^{2 \sqrt{\xi_H} \chi/M_P},  \\
\epsilon &\simeq&128 \xi_{H}  e^{-4 \sqrt{\xi_H} \chi/M_P} \simeq \frac{1}{8 N^{2} \xi_{H}}, \\
\eta &\simeq& -32 \xi_{H}  e^{-2 \sqrt{\xi_H} \chi/M_P} \simeq-\frac{1}{N}.
\eea
Thus, we obtain the approximate results for the inflationary observables,
\begin{align}
A_{s} \simeq \frac{N^{2}}{12 \pi^{2}} \frac{\lambda_H}{ \xi_{H}}, \quad n_{s} \simeq 1-\frac{2}{N}, \quad r \simeq \frac{2}{N^{2} \xi_{H}}. \label{simpinf}    
\end{align}
Taking the observed value $A_s=2.1\times 10^{-9}$ with $N= 50-60$, we have the following relation
\begin{align}
 \xi_H =1.0\times 10^{10}\lambda_H,   \label{cmb}
\end{align}
in comparison to the case in the metric formulation for Higgs inflation which requires $\xi_H\sim 10^4\sqrt{\lambda_H}$. 

We also get the spectral index, $n_s=0.960-0.967$, for $N=50-60$, which is consistent with the Planck data, $n_s=0.967\pm 0.0037$ \cite{planck}. Thus, the inflationary predictions in our model coincide with those in the Palatini formulation for Higgs inflation \cite{Bauer:2008zj,Takahashi:2018brt,Jinno:2019und}. Namely, the non-minimal coupling $\xi_H$ is larger than that of the original Higgs inflation for a sizable $\lambda_H$, resulting in quite a tiny tensor-to-scalar ratio $r$. However, even for a large $\xi_H$, the cutoff scale does not fall below the typical energy scales of inflation and reheating~\cite{Mikura:2021clt}.

 We remark that  from the Einstein-frame Lagrangian in eq.~(\ref{Einsteinsimple}), the Weyl gauge field $X_\mu$ gets mass during inflation, as follows, 
 \bea
 m^2_X&=&\frac{g^2_X h^2}{1+\xi_H h^2/M^2_P} \nonumber \\
& \simeq& \frac{g^2_X M^2_P}{\xi_H} \Big(1+e^{-\sqrt{\xi_H}\chi/M_P}\Big)^{-1}, 
\eea
which is sufficiently larger than the Hubble scale, $H^2_I\simeq \lambda_H M^2_P/(4\xi^2_H)$, for $g^2_X\gg \lambda_H/(4\xi_H)$. 
Therefore, we can safely ignore the dynamics of the Weyl gauge field $X_\mu$ during inflation.

\section{General couplings for two Weyl gauge fields}

We generalize the discussion in the previous section with extra Weyl-covariant kinetic terms for the dilaton and the Higgs fields and discuss the unitarity scale and the inflationary predictions.

\subsection{General Lagrangian in Weyl gravity}

Introducing the extra parameters, $r_\phi$ and $r_H$, and using the results in Appendix B, we consider the most general Lagrangian in Weyl gravity with two Weyl gauge fields and the Ricci scalar given by eq.~(\ref{Ricci}), as follows,
\bea
\frac{{\cal L}_G}{\sqrt{-g}} &=& -\frac{1}{2}(\xi_\phi \phi^2+\xi_H |H|^2) {\tilde R}+ 3\xi_\phi  (r_\phi-1) (D_\mu \phi)^2 +6\xi_H (r_H-1) |D_\mu H|^2  \nonumber \\
 &&-\frac{1}{4} w_{\mu\nu} w^{\mu\nu}- \frac{1}{4} X_{\mu\nu} X^{\mu\nu}  +(1-6\xi_H (r_H-1)) |D'_\mu H|^2 - V(H,\phi). \label{weylinvg00}
\eea
Then, we expand the above Lagrangian by using eq.~(\ref{Ricci}), as follows,
\bea
\frac{{\cal L}_G}{\sqrt{-g}} &=& -\frac{1}{2} \xi_\phi \bigg(\phi^2 R +6(\partial_\mu\phi)^2 - 6r_\phi (D_\mu \phi)^2 \bigg) -\frac{1}{4} w_{\mu\nu} w^{\mu\nu}  \nonumber \\
&&-\xi_H \bigg(|H|^2 R+ 6 |\partial_\mu H|^2-6 r_H  |D_\mu H|^2 \bigg) \nonumber \\
 &&- \frac{1}{4} X_{\mu\nu} X^{\mu\nu}  +(1-6\xi_H(r_H-1)) |D'_\mu H|^2 - V(H,\phi). \label{weylinvg0}
\eea
Here, we note that the net Higgs kinetic term is normalized to be canonical in Jordan frame.

Following a similar step in the Appendix B as in the case with $r_\phi=r_H=1$, we obtain the Einstein-frame Lagrangian with gauge fixing, $\langle\phi^2\rangle=M^2_P/\xi_\phi$, as follows, 
\bea
\frac{{\cal L}_E}{\sqrt{-g_E}} &=&-\frac{M^2_P}{2} R+6\xi_H(r_H-1)\frac{|\partial_\mu H|^2}{\Omega}+ (1-6\xi_H(r_H-1)) \frac{|D'_\mu H|^2}{\Omega} \nonumber \\
&&+\frac{3\xi_H^2}{M_P^2\Omega^2}\frac{r_{\phi}-r^2_H+2r_H(1-r_H)\xi_H|H|^2/M_P^2}{r_{\phi}+2r_H\xi_H |H|^2/M_P^2}\left(\partial_{\mu}|H|^{2}\right)^2  \nonumber \\
&&- \frac{V(H)}{\Omega^2}- \frac{1}{4} X_{\mu\nu} X^{\mu\nu} -\frac{1}{4} {\tilde w}_{\mu\nu} {\tilde w}^{\mu\nu} + \frac{1}{2}(m^2_w + g^2_w r_H K_H)  \Omega^{-1} {\tilde w}_\mu {\tilde w}^\mu \label{genWeyl0}
\eea
where ${\tilde w}_\mu$ is the redefined heavy Weyl gauge field, given by
\bea
{\tilde w}_\mu = w_\mu-\frac{ 1}{2g_w} \partial_\mu \ln (m^2_w+ g^2_w r_H K_H),
\eea
with $K_H=12\xi_H |H|^2$, 
and the mass of the heavy Weyl gauge field becomes $m^2_w=6 r_\phi g^2_w M^2_P$. 
Then, the heavy Weyl gauge field ${\tilde w}_\mu$ is generically not decoupled from the Higgs field, except for $r_\phi=r_H$.

We now discuss the unitarity scale in the presence of the general gravitational couplings. To this, we focus on the non-canonical Higgs kinetic terms, as follows,
\bea
\frac{\mathcal{L}_{\rm{kin}}}{\sqrt{-g_E}} &=&  \frac{ |\partial_\mu H|^2}{1+2\xi_H |H|^2/M_P^2} \nonumber \\
&&+\frac{3\xi_H^2}{M_P^2}\frac{r_{\phi}-r^2_H+2r_H (1-r_H) \xi_H |H|^2/M_P^2}{(1+2\xi_H |H|^2/M_P^2)^2(r_{\phi}+2r_H \xi_H |H|^2/M_P^2)}\cdot (\partial_\mu |H|^2)^2.    \label{genHkin}
\eea
Therefore, the cutoff scale depends on various combinations of $\xi_H$, $r_\phi$ and $r_H$, but it can be generically much larger than the one in Higgs inflation. For instance, we obtain the leading dimension-6 operator for the Higgs, 
\bea
\frac{\mathcal{L}_{\rm{kin}}}{\sqrt{-g_E}} \supset && -\frac{2\xi_H}{M^2_P} |H|^2 |\partial_\mu H|^2 +\frac{3\xi^2_H}{M^2_P} \Big(1-\frac{r^2_H}{r_\phi} \Big)(\partial_\mu |H|^2)^2 \nonumber \\
&&=\frac{1}{M^2_P} \Big(3\xi^2_H \Big(1-\frac{r^2_H}{r_\phi} \Big) +\xi_H \Big)(\partial_\mu |H|^2)^2 +\cdots
\eea
where integration by parts are made and the equation of motion for the Higgs fields is used in the second line.
Thus, the unitarity scale identified from the  leading dimension-6 operator is
\bea
\Lambda_1= \frac{M_P}{ \Big|3\xi^2_H \Big(1-\frac{r^2_H}{r_\phi} \Big) +\xi_H \Big|^{1/2}},
\eea
which can be much larger than the one in Higgs inflation, depending on $\xi_H, r_\phi$ and $r_H$.
On the other hand, even higher dimensional derivative terms and the higher dimensional terms in the scalar potential lead to the cutoff scale, $\Lambda=M_P/\sqrt{\xi_H}$.

For $r_\phi=r_H$,  the Einstein-frame Lagrangian becomes
\bea
\frac{{\cal L}_E}{\sqrt{-g_E}} &=&-\frac{M^2_P}{2} R+6\xi_H(r_H-1) \frac{|\partial_\mu H|^2}{\Omega}+(1-6\xi_H(r_H-1)) \frac{|D'_\mu H|^2}{\Omega}   \nonumber \\
&& +\frac{3\xi^2_H(1-r_H)}{\Omega^2 M_P^2} (\partial_\mu |H|^2)^2 - \frac{V(H)} {\Omega^2} \nonumber \\
&&- \frac{1}{4} X_{\mu\nu} X^{\mu\nu}-\frac{1}{4} {\tilde w}_{\mu\nu} {\tilde w}^{\mu\nu} + \frac{1}{2} m^2_w  {\tilde w}_\mu {\tilde w}^\mu,
\label{Einsteinsimple2}
\eea
so the heavy Weyl gauge field ${\tilde w}_\mu$ is decoupled from the Higgs fields in Einstein frame.
In this case, we find that the unitarity scale is given by $\Lambda_1=M_P/\sqrt{3 \xi^2_H(1-r_H)+\xi_H}$ from the leading dimension-6 operator and $\Lambda=M_P/\sqrt{\xi_H}$ from the other higher dimensional terms.

\subsection{Inflation with $r_{\phi}=r_H$}

For $r_\phi=r_H$,  the heavy Weyl gauge field is decoupled from the Higgs field and the Higgs kinetic term in eq.~(\ref{genHkin}) becomes simplified. For $r_\phi\neq r_H$, the Weyl gauge field is not decoupled from the Higgs in eq.~(\ref{genWeyl0}), but a slow-roll inflation can be maintained even in this case, as far as the Weyl gauge field is sufficiently heavier than the Hubble scale during inflation. For simplicity  in the following discussion, we choose $r_\phi=r_H$ for the inflation with a varying unitarity scale depending on $r_H$. 

In the following, we take  $r_\phi=r_H$, for which the relevant Einstein-frame Lagrangian for inflation is simplified to
\bea
\frac{{\cal L}_{E,{\rm inf}}}{\sqrt{-g_E}} =-\frac{M_P^2}{2} R+\frac{1}{2\Omega^2}\, \Big(1+\xi_H\left(1+3\xi_H(1-r_H)\right) h^2/M_P^2\Big) (\partial_\mu h)^2 - \frac{V(h)}{\Omega^2} \label{generalinf}
\eea
with $\Omega=1+\xi_H h^2/M_P^2$. The parameter $r_H$ must satisfy $r_H>0$, for the Weyl gauge field mass to be positive. Also, to avoid a ghost mode in the large field region with $\xi_H^2h^2\gg M_P^2$, we consider a case with $0<r_H<1$. In the following discussion, we set $M_P=1$ for notational simplicity.

In order to see the modification for $r_H\neq 1$, we first introduce the field redefinition by
\bea
\xi_H h^2 = e^{2\sqrt{\xi_H}{\hat \chi}}. 
\eea
Then, we can rewrite eq.~(\ref{generalinf}) in terms of $\hat\chi$ as
\bea
\frac{{\cal L}_{E,{\rm inf}}}{\sqrt{-g_E}} = -\frac{M_P^2}{2} R+\frac{1}{2} \frac{(1+3\xi_H(1-r_H)+e^{-2\sqrt{\xi_H} {\hat \chi}})}{\left(1+ e^{-2\sqrt{\xi_H} {\hat \chi}}\right)^2} (\partial_\mu {\hat\chi})^2 -\frac{\lambda_H}{4\xi^2_H} \Big( 1+ e^{-2\sqrt{\xi_H} {\hat \chi}}\Big)^{-2}.
\eea
We can make a further redefinition of the inflaton by
\begin{align}
\nonumber \chi=\ &\sqrt{1/\xi_H+3(1-r_H)}{\rm{Arctanh}}\sqrt{\frac{1+3\xi_H(1-r_H)+e^{-2\sqrt{\xi_H} {\hat \chi}}}{1+3\xi_H(1-r_H)}}\\
&-\sqrt{3(1-r_H)}{\rm{Arccoth}}\sqrt{\frac{3\xi_H(1-r_H)}{1+3\xi_H(1-r_H)+e^{-2\sqrt{\xi_H} {\hat \chi}}}}.    
\end{align}
But, this is not of invertible form for the canonical inflaton $\chi$. 
Thus, instead we compute the slow-roll parameters and the number of efoldings in terms of $\hat\chi$ as below.

First, for $e^{-2\sqrt{\xi_H}{\hat \chi}}\ll 1$, we obtain the slow-roll parameters as
\bea
\epsilon&=&\frac{1}{2V} \Big(\frac{d{\hat \chi}}{d\chi}\Big)^2 \Big(\frac{d V}{d{\hat \chi}}\Big)^2 
 \simeq  \frac{8\xi_H\, e^{-4\sqrt{\xi_H}{\hat \chi}}}{1+3\xi_H(1-r_H)}, \\
 \eta &=&\frac{d{\hat \chi}}{d\chi} \frac{d}{d{\hat\chi}}\Big(\frac{d V}{d\chi}\Big) \simeq -\frac{8\xi_H\, e^{-2\sqrt{\xi_H}{\hat \chi}}}{1+3\xi_H(1-r_H)},
\eea
and the number of efoldings is
\bea
N&=&\int ^{\chi}_{\chi_e} \frac{d \hat{\chi}}{d \ln V / d \hat{\chi}}\left(\frac{d \chi}{d \hat{\chi}}\right)^{2} \nonumber \\
&\simeq&\frac{e^{2\sqrt{\xi_H}{\hat \chi}}}{8\xi_H} \Big(1+3\xi_H(1-r_H) -6\xi^{3/2}_H(1-r_H)\,{\hat\chi}\, e^{-2\sqrt{\xi_H}{\hat \chi}} \Big). \label{Nefoldgen}
\eea
Then, we can get an approximate solution  for $\hat{\chi}$ to the above equation, as follows,
\begin{align}
e^{-2\sqrt{\xi_H}{\hat \chi}}  \simeq  \frac{1+3\xi_H(1-r_H)}{8\xi_HN}\left[1-\frac{3}{8N}(1-r_H)\ln \frac{8\xi_HN}{1+3\xi_H(1-r_H)}\right].
\end{align}

As a result, we can rewrite the slow-roll parameters in terms of the number efoldings, as follows,
\bea
\epsilon&\simeq & \frac{1+3\xi_H(1-r_H)}{8\xi_H N^2}\,\left[1-\frac{3}{8N}(1-r_H)\ln \frac{8\xi_HN}{1+3\xi_H(1-r_H)}\right]^2 , \\
\eta &\simeq & -\frac{1}{N}\,  \left[1-\frac{3}{8N}(1-r_H)\ln \frac{8\xi_HN}{1+3\xi_H(1-r_H)}\right].
\eea
Consequently, we get the general expressions for the inflationary observables as
 \bea
 A_s &\simeq & \frac{N^2}{12\pi^2} \frac{\lambda_H}{\xi_H}\frac{1}{1+3\xi_H(1-r_H)} \, \left[1-\frac{3}{8N}(1-r_H)\ln \frac{8\xi_HN}{1+3\xi_H(1-r_H)}\right]^{-2},\\
 n_s&\simeq& 1- \frac{2}{N}\,  \left[1-\frac{3}{8N}(1-r_H)\ln \frac{8\xi_HN}{1+3\xi_H(1-r_H)}\right], \label{n_s}\\
 r&\simeq & \frac{2(1+3\xi_H(1-r_H))}{\xi_H N^2}\,\left[1-\frac{3}{8N}(1-r_H)\ln \frac{8\xi_HN}{1+3\xi_H(1-r_H)}\right]^2.\label{r}
 \eea
Then, for $r_H=1$, we recover the results in eq.~(\ref{simpinf}). Otherwise, all the inflationary observables are different from the case with $r_H=1$. In particular, the spectral index can be corrected sizably. For instance, from $\xi_H= 10^{10} \lambda_H$, $\lambda_H=0.01$ and $N=50$, we get $\frac{3(r_H-1)}{4N} \,\ln(8\xi_H N)= 0.4 (r_H-1)$, so the deviation in the spectral index can be significant.

Moreover, for $r_H<1$, the CMB normalization determines the non-minimal coupling approximately by
\bea
\xi_H \simeq \frac{1}{6(1-r_H)} \Big(\sqrt{1+N^2(1-r_H) \lambda_H/(\pi^2 A_s)} -1\Big). \label{cmb2}
\eea
For $N^2(1-r_H) \lambda_H/(\pi^2 A_s)\ll 1$, we recover the value of $\xi_H$ in eq.~(\ref{cmb}) with $r_H=1$; for $N^2(1-r_H) \lambda_H/(\pi^2 A_s)\gg 1$, we get $\xi_H\simeq (N/6\pi)\sqrt{\lambda_H/(A_s(1-r_H))}\simeq 6\times 10^3/\sqrt{1-r_H}$ for $\lambda_H=0.01$ and $N=50$. 

\begin{figure}[t]

  \begin{center}
   \includegraphics[width=120mm]{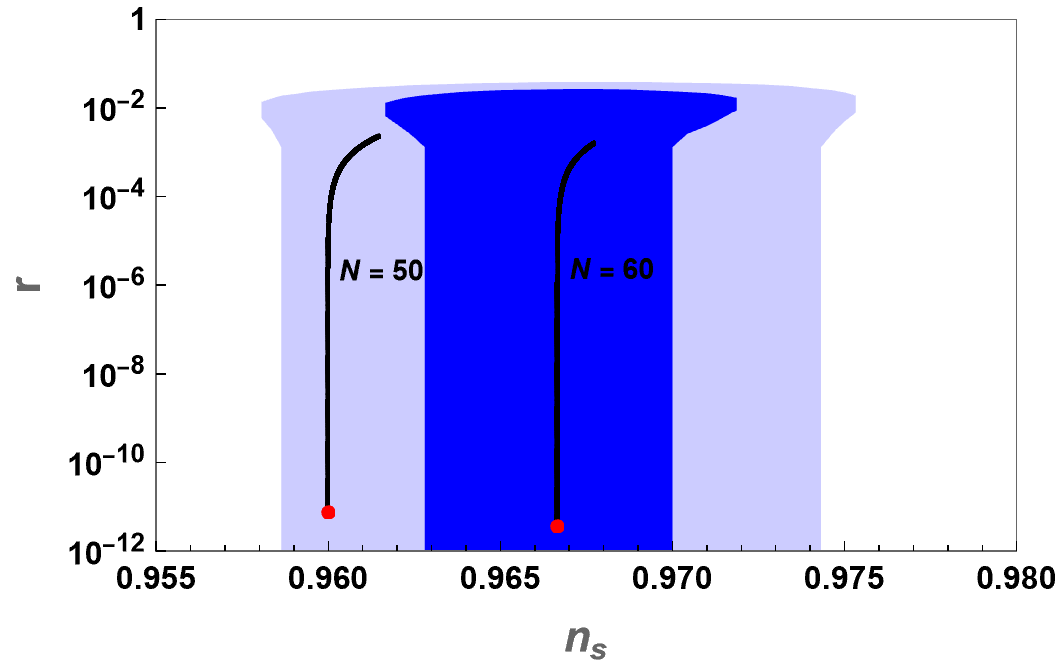}
  \end{center}
\caption{The spectral index $n_s$ and the tensor-to-scalar ratio $r$ as a function of $r_H$. The bounds on $n_s$ and $r$, from the combined results of Planck, WMAP, BICEP and Keck data, are shown in dark and light blue regions, within $1\sigma$ and $2\sigma$ errors, respectively. The red points correspond to $r_H=1$ for $N=50, 60$, respectively.}
  \label{ns_r}
 
\end{figure}

In Fig.~\ref{ns_r}, we show the predicted values for the spectral index $n_s$ and the tensor-to-scalar ratio $r$ by fixing $N=50,60$ and $\lambda_H=0.01$ but varying $r_H$ in the range of $0<r_H<1$. We set $\xi_H$ from the CMB normalization, and imposed the bounds on $n_s$ and $r$, for Planck data combined with WMAP, BICEP and Keck data \cite{bicep}.  The dark blue and light blue regions are within $1\sigma$ and $2\sigma$ errors. The red dots are for $r_H=1$ with $N=50, 60$, respectively, which is equivalent to the minimal Weyl gravity or the Palatini formulation for Higgs inflation.  But, for $r_H\neq 1$, our model connects between the Palatini formulation for Higgs inflation and a Higgs-like inflation continuously. There was a similar construction of interpolating between the original Higgs inflation and its Palatini formulation in the context of Einstein-Cartan gravity \cite{cartan}.

We also remark that  from the Einstein-frame Lagrangian in eq.~(\ref{Einsteinsimple2}), the Weyl gauge field $X_\mu$ gets mass during inflation, as follows, 
 \bea
 m^2_X&=&\frac{a_H g^2_X h^2}{1+\xi_H h^2/M^2_P} \nonumber \\
& \simeq& \frac{a_H g^2_X M^2_P}{\xi_H} \Big(1+e^{-\sqrt{\xi_H}{\hat\chi}/M_P}\Big)^{-1}, 
\eea
with $a_H=1-6\xi_H(r_H-1)$. 
Thus, the Weyl gauge field mass is sufficiently larger than the Hubble scale for $a_H g^2_X\gg \lambda_H/\xi_H$ as in the case with $r_H=1$ in the previous subsection, so we can safely ignore the dynamics of the Weyl gauge field $X_\mu$ in the case with general Weyl invariant terms.

\section{Light Weyl gauge field and Higgs physics}

The Weyl gauge field $w_\mu$ coupled to the dilaton $\phi$ gets a large mass after the Weyl symmetry is broken spontaneously. 
On the other hand, another Weyl gauge field $X_\mu$ couples only to the Higgs fields and extra light scalars, so it receives a mass at low energy.  As the Higgs fields couple to the light Weyl gauge field directly via the covariant kinetic term,  we discuss some interesting implications for Higgs physics and new resonance searches at colliders.

\subsection{Higgs mechanism for light Weyl gauge field}

In our model, the extra Weyl gauge field $X_\mu$ can be light, as far as it couples to an extra light scalar $s$, with the covariant kinetic term in Jordan frame, as follows,
\bea
\frac{\Delta {\cal L}}{\sqrt{-g}} = \frac{1}{2}(D'_\mu s)^2 \label{singletkin}
\eea
with  $D'_\mu s=(\partial_\mu -g_X X_\mu)s$. We note that the extra light scalar $s$ also couples to the heavy Weyl gauge field $w_\mu$, being consistent with the Weyl symmetry, for instance, through $D_\mu s=(\partial_\mu -c\,g_w w_\mu-(1-c)g_X X_\mu)s$, with $c$ being constant. Then, there appears a mixing between $w_\mu$ and $X_\mu$ in this case. But, for $g_X \langle s\rangle\ll g_w \langle\phi\rangle=g_w M_P/\sqrt{\xi_\phi}$, we can safely decouple the heavy Weyl gauge field $w_\mu$, without changing our conclusions below.

Then, from eq.~(\ref{singletkin}), we get the leading interactions between the light Weyl gauge field and the Higgs and singlet fields in Einstein frame as
\bea
{\cal L}_{X,{\rm int}} &=& a_H \Big(-g_X X_\mu\partial^\mu |H|^2+ g^2_X X_\mu X^\mu |H|^2 \Big)  \nonumber \\
&&   -\frac{1}{2}g_X X_\mu\partial^\mu s^2+ \frac{1}{2}g^2_X X_\mu X^\mu s^2,  \label{Xint}
\eea
with $a_H\equiv 1-6\xi_H(r_H-1)$.
Here, we assume that the extra singlet scalar $s$ does not change  the previous discussion on inflation for $s=0$ during inflation, but it is crucial  for the Higgs mechanism for the light Weyl gauge field. 

After the Higgs and  the singlet get VEVs, we expand them by $H=(0,v+h)^T/\sqrt{2}$, and $s=v_s+{\tilde s}$.
Then, the light Weyl gauge field $X_\mu$ gets a nonzero mass by
\bea
m^2_X=g^2_X (a_H v^2 + v^2_s). 
\eea
Moreover, there appears a mixing between the Weyl gauge field and the Higgs fields in eq.~(\ref{Xint}), which can be eliminated by a gauge fixing term for the Weyl gauge field,
\bea
{\cal L}_{gf} &=& -\frac{1}{2\zeta} \bigg(\partial_\mu X^\mu +\zeta  g_X (a_H v h + v_s{\tilde s}) \bigg)^2, \label{gaugefix}
\eea
with $\zeta$ being the gauge fixing parameter. Thus,  the would-be Goldstone, $G_X\sim a_H v h+  v_s {\tilde s}$, is eaten by the Weyl gauge field $X_\mu$, and  the orthogonal combination, $h_{\rm SM}\sim  v_s h-a_H v {\tilde s}$, is identified as the SM Higgs boson.  From the gauge fixing term in eq.~(\ref{gaugefix}), the would-be Goldstone has mass, $m^2_{G_X}=\zeta g^2_X ((a_H v)^2+ v_s^2)$, so the Goldstone $G_X$ would be decoupled in unitary gauge with $\zeta\to \infty$ as usual in spontaneously broken gauge theories.

We remark that in order to keep the would-be Goldstone boson massless in the scalar potential, we need to take the alignment limit for the mass mixing between Higgs and singlet scalars. To this, we need to set the quartic potential for them to $\lambda_H\big(|H|^2+\frac{1}{2a_H}s^2\big)^2$. Then, there would be a  generic fine-tuning for the mixing quartic coupling between the singlet and the Higgs, but the aforementioned quartic coupling is protected by the $SO(5)$ symmetry in the limit of $a_H=1$ or $r_H=1$, for which the unitarity scale becomes maximized to $M_P/\sqrt{\xi_H}$. Then, during inflation, the effective mass for the singlet scalar becomes $m^2_s\simeq \lambda_H M^2_P/(2\xi_H a_H)$ for $|H|^2\gtrsim M^2_P/(2\xi_H)$, which is much larger than the Hubble scale during inflation, $H^2_I\sim \lambda_H M^2_P/(12\xi^2_H)$, for a large $\xi_H$. Therefore, we can safely ignore the dynamics of the singlet scalar $s$ during inflation.

\subsection{Weyl gauge field interactions}

As we discussed with the gauge-fixing Lagrangian in the previous subsection, we introduce a mixing angle between the Higgs and singlet scalars by
\bea
\left(\begin{array}{c} h \\ {\tilde s} \end{array}\right) = \left( \begin{array}{cc}  \cos\theta & \sin\theta \\  -\sin\theta & \cos\theta \end{array}\right) \left(\begin{array}{c} h_{\rm SM} \\ G_X \end{array}\right).
\eea
with $\tan\theta=(a_H v)/v_s$. 
Since the SM particles couple only to $h$, the SM Higgs couplings for the physical scalar, $h_{\rm SM}$, are modified as compared to the SM. Thus, the decays and production channels for the SM Higgs are modified by the mixing angle. Therefore, in order to be consistent with Higgs data, we need to take the mixing angle sufficiently small, for instance, $\sin\theta\simeq (a_Hv)/v_s\lesssim 0.3(0.03)$ at $10(1)\%$ level.

Taking the general covariant gauge with the gauge fixing term for the Weyl gauge field in eq.~(\ref{gaugefix}), we can reduce the interaction terms of the Weyl gauge field to the would-be Goldstone $G_X$ and the SM Higgs boson $h_{\rm SM}$, in the following,
\bea
{\cal L}_{X,{\rm int}}&=&-a_Hg_X X_\mu h \partial^\mu h+ \frac{1}{2} a_H g^2_X X_\mu X^\mu (h^2+2 vh) \nonumber \\
&&- g_X X_\mu {\tilde s}\partial^\mu {\tilde s} + \frac{1}{2}  g^2_X X_\mu X^\mu ({\tilde s}^2+2 v_s {\tilde s}) \nonumber \\
&=& -g_X (a_H \cos^2\theta+\sin^2\theta) X_\mu h_{\rm SM} \partial^\mu h_{\rm SM}  \nonumber \\
&&-g_X (a_H-1)\cos\theta\sin\theta X_{\mu}(G_X\partial^\mu h_{\rm SM}+h_{\rm SM} \partial^\mu G_X) \nonumber \\
&&-g_X (a_H \sin^2\theta+\cos^2\theta) X_\mu G_X \partial^\mu G_X \nonumber \\
 &&+g^2_X \sqrt{(a_H v)^2+v_s^2}\, X_\mu X^\mu G_X +\frac{1}{2} g^2_X (a_H\cos^2\theta+ \sin^2\theta) X_\mu X^\mu  h^2_{\rm SM} \nonumber \\
&&  +g^2_X (a_H-1)\cos\theta\sin\theta\,  X_\mu X^\mu G_X  h_{\rm SM} \nonumber \\
&&+\frac{1}{2} g^2_X (a_H\sin^2\theta+\cos^2\theta) X_\mu X^\mu  G^2_X. 
  \label{Xint2}
\eea
Here, we note that the perturbativity conditions on the Weyl gauge field couplings are given by 
\bea
g_X<1, \quad a_H g^2_X <1.
\eea

In unitary gauge, the would-be Goldstone is decoupled, so we only have to consider the Weyl gauge field couplings for the SM Higgs, $X_\mu h_{\rm SM} \partial^\mu h_{\rm SM}$ and $X_\mu X^\mu h^2_{\rm SM}$. But, the derivative couplings of the Weyl gauge field can be written as $X_\mu h_{\rm SM} \partial^\mu h_{\rm SM}=-\frac{1}{2}\partial_\mu X^\mu h^2_{\rm SM}$, etc, up to total derivative terms, so there is no decay of the on-shell Weyl gauge field because of $\partial_\mu X^\mu=0$.
Moreover, we note that there is no linear Higgs coupling to the Weyl gauge field, so there is no additional decay of the SM Higgs boson. 
Since there is no direct coupling between the Weyl gauge field and the other SM particles (fermions and gauge bosons), it is challenging to test the Weyl gauge field at current collider experiments. 

We remark the effects of a gauge kinetic mixing for the Weyl gauge field for testing the Weyl gauge field models. 
In the presence of a gauge kinetic mixing between the Weyl gauge field and the hypercharge gauge boson $B_\mu$ in the following form,
\bea
{\cal L}_{\rm gmix} = -\frac{1}{2} \sin\xi \, X_{\mu\nu} B^{\mu\nu}
\eea 
with $B_{\mu\nu}$ being the gauge field strength for the hypercharge gauge boson, we need to diagonalize the gauge kinetic terms and the mass matrix for neutral gauge bosons simultaneously. As a result, the electroweak neutral gauge bosons and the Weyl gauge field are mixed \cite{Z3,VSIMP} by
\begin{equation}
\left(
\begin{array}{c}
 B_\mu \\
 W_{3\mu} \\
 X_\mu
\end{array}
\right)=\left(
\begin{array}{ccc}
 c_W & t_{\xi } s_{\zeta }-s_W c_{\zeta } & -s_W s_{\zeta }-t_{\xi } c_{\zeta }  \\
 s_W & c_W c_{\zeta } & c_W s_{\zeta }  \\
 0 & -s_{\zeta }/c_{\xi } & c_{\zeta }/c_{\xi } 
\end{array}
\right) \left(
\begin{array}{c}
 \tilde{A}_\mu \\
 \tilde{Z}_\mu \\
 \tilde{X}_\mu 
\end{array}
\right) \label{gaugerot}
\end{equation}
Here, $c_\xi\equiv \cos\xi$, $t_\xi\equiv \tan\xi$,  $c_W\equiv \cos\theta_W$,  $s_W\equiv \sin\theta_W$, and $\zeta$ is the mixing angle between $Z$ and $X$ bosons, given \cite{Z3,VSIMP} by
\bea
\tan(2\zeta)=\frac{m^2_Z s_W \sin(2\xi)}{m^2_X-m^2_Z (c^2_{\xi}-s^2_W s^2_\xi)}.
\eea
In the limit of $m^2_X\ll m^2_Z$ (or $m^2_X\gg m^2_Z$) and $|\xi|\ll 1$, we obtain $\zeta\approx -s_W \xi$ (or $\zeta\approx \frac{m^2_Z}{m^2_X} \, s_W \xi$). 
Moreover, the mass eigenvalues for $Z$-like and $X$-like gauge bosons are given \cite{Z3,VSIMP} by
\bea
m^2_{1,2}=\frac{1}{2} \left[ m^2_Z (1+s^2_W t^2_\xi)+ m^2_X/c^2_\xi\pm \sqrt{\Big(m^2_Z(1+s^2_W t^2_\xi)+m^2_X/c^2_\xi\Big)^2-4m^2_Z m^2_X/c^2_\xi}\right].
\eea

We first recall that the current interactions in the interaction basis are given by
\bea
{\cal L}_{\rm EM/NC}= g_X X_\mu J^\mu_X+e(s_W W_{3\mu}+c_W B_\mu) J^\mu_{\rm EM} +\frac{e}{2s_W c_W}\,(c_W W_{3\mu}- s_W B_\mu )J^\mu_Z
\eea
where $J_X^\mu$ are the Weyl current given by $J_X^\mu=-a_H \partial_\mu |H|^2+\frac{1}{2} \partial_\mu s^2$, and $ J^\mu_{\rm EM} $ and $J^\mu_Z$ are electromagnetic and neutral currents. 
Then, using eq.~(\ref{gaugerot}), we can approximate the above current interactions in the basis of mass eigenstates \cite{Z3,VSIMP}, for $\varepsilon\equiv c_W t_\xi\approx c_W\xi\ll 1$, as
 \bea
{\cal L}_{\rm EM/NC}\simeq e \tilde{A}_\mu J^\mu_{\rm EM} + \tilde{Z}_\mu  \bigg[ \frac{e}{2s_Wc_W} J^\mu_Z + \varepsilon g_X t_W  J^\mu_X \bigg] +\tilde{X}_\mu \bigg[ g_X  J_X^\mu -e \varepsilon  J_\text{EM}^\mu   \bigg]. 
\eea
Therefore, we find that the redefined Weyl gauge field $\tilde{X}_\mu$ couples to the electromagnetic current $ J_\text{EM}^\mu$ too, so it is possible to produce the Weyl gauge field directly at the LHC and other collider experiments. Consequently, we can test the Weyl gauge field models by the interplay of standard decay modes into a pair of SM particles (apart from Higgs) appearing in non-Weyl $Z'$ models and the Weyl current interaction.

\section{Conclusions}

We considered the embedding of Higgs inflation with a non-minimal coupling into the Weyl gravity where the unitarity problem of the original Higgs inflation is less severe, thanks to the heavy Weyl gauge field coupled to the Higgs fields. When the couplings of the heavy Weyl gauge field are absorbed into the non-minimal couplings to the Ricci curvature scalar in Weyl gravity, we found that the resultant model for Higgs inflation is the same as in the Palatini formulation for Higgs inflation.  The covariant kinetic term for the Higgs fields with the second light Weyl gauge field is necessary for a successful Higgs inflation with a large non-minimal coupling. Thus, the crucial difference of our model from the Palatini formulation for Higgs inflation is that there is a light Weyl gauge field coupled to the Higgs fields.

We also generalized the unitarization of Higgs inflation with Weyl gauge fields in the presence of general covariant kinetic terms for the dilaton and the Higgs fields. In this case, we realized a successful Higgs inflation, interpolating between the Palatini formulation for Higgs inflation and a Higgs-like inflation. However, due to the unitarity problem, the region of the parameter space close to the Palatini formulation for Higgs inflation is favored. 

We also showed that the light Weyl gauge field gets a small mass due to the VEV of an extra singlet scalar and the Goldstone boson associated with the light Weyl gauge field is a mixture of the Higgs and extra singlet scalars. Therefore, we found that the mixing between the Higgs and extra singlet scalars can be constrained by Higgs data. Furthermore, in the presence of a gauge kinetic mixing for the light Weyl gauge field, there are interesting signatures for $Z'$-like resonances at the LHC, with the direct couplings of the light Weyl gauge field to the SM Higgs bosons, unlike in usual $Z'$ models.

\section*{Acknowledgments}

The work is supported in part by Basic Science Research Program through the National
Research Foundation of Korea (NRF) funded by the Ministry of Education, Science and
Technology (NRF-2022R1A2C2003567 and NRF-2021R1A4A2001897).

\def\theequation{A.\arabic{equation}}

\setcounter{equation}{0}

\vskip0.8cm
\noindent
{\Large \bf Appendix A: Comparison to unitarization with singlet scalars}

\underline{\bf Linear sigma models}

We consider an induced gravity model for the sigma field $\sigma$, with the following Jordan frame Lagrangian \cite{unitarity,unitarity2},
\bea
\frac{{\cal L}_\sigma}{\sqrt{-g}}= -\frac{1}{2} \sigma^2 R  +\frac{1}{2} (\partial_\mu\sigma)^2 - \frac{1}{4} \lambda_\sigma \bigg(\sigma^2-M^2_P-2\xi_H |H|^2 \bigg)^2.  \label{sigma}
\eea

Then, after integrating out the sigma field with its equation of motion,
\bea
\sigma^2=M^2_P+2\xi_H |H|^2,
\eea
we obtain the effective non-minimal coupling for the Higgs from the one for the sigma field in eq.~(\ref{sigma}).

From eq.~(\ref{sigma}), the effective Higgs quartic coupling is related to the running quartic couplings by
\bea
\lambda_{\rm eff}=\lambda_H-\lambda_\sigma \xi^2_H
\eea
where $\lambda_\sigma \xi^2_H<1$ should be taken from perturbativity.

\underline{\bf Starobinsky model}

We can add an $R^2$ term in Higgs inflation by
\bea
{\cal L}_{R2} =\sqrt{-g} \,\alpha R^2,
\eea
which is dual to the scalaron Lagrangian,
\bea
{\cal L}_{R2}= -2\alpha \chi R -\alpha \chi^2.
\eea
Then, after the field redefinitions,
\bea
g_{\mu\nu}\to \Omega^2 g_{\mu\nu}, \quad \chi \to \Omega^2\chi,\quad  H\to \Omega H,
\eea
with $\Omega^{-2}= \big(1+\frac{\sigma}{\sqrt{6}} \big)^2$,
and
\bea
\Omega^{-2} +2\xi_H|H|^2 +4\alpha\chi= 1-\frac{1}{3} |H|^2 -\frac{1}{6}\sigma^2,
\eea
it was shown that the Higgs-$R^2$ Lagrangian can be recast into a linear sigma model type \cite{ema,general,HiggsR2susy,HiggsR2}, 
\bea
\frac{{\cal L}_{H+R2} }{\sqrt{-g}}&=& -\frac{1}{2} \bigg(1-\frac{1}{6}\sigma^2-\frac{1}{3}|H|^2 \bigg)R +\frac{1}{2}(\partial_\mu\sigma)^2 +|\partial_\mu H|^2 \nonumber \\
&&- \frac{1}{144\alpha} \bigg[ \Big(\sigma+\frac{\sqrt{6}}{2}\Big)^2+6\Big(\xi_H+\frac{1}{6}\Big)|H|^2-\frac{3}{2}\bigg]^2.
\eea
As a result, there is no unitarity problem below the Planck scale, as far as the following perturbativity conditions are fulfilled,
\bea
\frac{1}{\alpha} \Big(\xi_H+\frac{1}{6}\Big)<1, \quad \frac{1}{\alpha} \Big(\xi_H+\frac{1}{6}\Big)^2<1.
\eea

In this case, there is a similar shift in the effective Higgs quartic coupling by
\bea
\lambda_{\rm eff} = \lambda_H-\frac{1}{4\alpha}  \Big(\xi_H+\frac{1}{6}\Big)^2.
\eea

\underline{\bf Singlet scalar with a triple coupling}

We consider a real singlet scalar $S$, with the following Lagrangian \cite{espinosa},
\bea
{\cal L}= \frac{1}{2} (\partial_\mu S)^2 -\frac{1}{2} m^2_S S^2- \mu S |H|^2.
\eea
Then, after integrating out the singlet scalar by 
\bea
S= -\frac{\mu}{m^2_S} |H|^2,
\eea
we get the effective interactions for the Higgs doublet as
\bea
{\cal L}_{\rm eff} =\frac{\mu^2}{2 m^4_S} (\partial_\mu |H|^2)^2 +\frac{\mu^2}{2m^2_S} |H|^4.
\eea 
Therefore, we obtain the desired non-canonical kinetic term for the Higgs with
\bea
\frac{\mu^2}{m^4_S} = \frac{3\xi^2_H}{M^2_P},
\eea
and there is a tree-level shift in the Higgs quartic coupling by
\bea
\lambda_{\rm eff}= \lambda_H- \frac{\mu^2}{2m^2_S}.
\eea

\def\theequation{B.\arabic{equation}}

\setcounter{equation}{0}

\vskip0.8cm
\noindent
{\Large \bf Appendix B: General Weyl-invariant Lagrangian}

Introducing extra parameters, $r_\phi$ and $r_H$, we consider the most general Lagrangian with Weyl invariance, as follows,
\bea
\frac{{\cal L}_G}{\sqrt{-g}} &=& -\frac{1}{2} \xi_\phi \bigg(\phi^2 R +6(\partial_\mu\phi)^2 - 6r_\phi (D_\mu \phi)^2 \bigg) -\frac{1}{4} w_{\mu\nu} w^{\mu\nu}  \nonumber \\
&&-\xi_H \bigg(|H|^2 R+ 6 |\partial_\mu H|^2-6 r_H  |D_\mu H|^2 \bigg) \nonumber \\
 &&- \frac{1}{4} X_{\mu\nu} X^{\mu\nu}  +(1-6\xi_H(r_H-1)) |D'_\mu H|^2 - V(H,\phi). \label{weylinvg}
\eea
Here, we chose the net Higgs kinetic term to be canonical in Jordan frame, without loss of generality. 
Then, fixing the gauge to $\langle\phi^2\rangle=M^2_P/\xi_\phi$, the gauge-fixed Lagrangian becomes
\bea
\frac{{\cal L}}{\sqrt{-g}} &=& -\frac{1}{2}  (M^2_P+2\xi_H |H|^2) R +6\xi_H(r_H-1)|\partial_\mu H|^2+ (1-6\xi_H(r_H-1)) |D'_\mu H|^2   \nonumber \\
&& -V(H)- \frac{1}{4} X_{\mu\nu} X^{\mu\nu}-\frac{1}{4} w_{\mu\nu} w^{\mu\nu} + \frac{1}{2}m^2_w w_\mu w^\mu \nonumber \\
&& -\frac{1}{2} r_H g_w w_\mu K^\mu+  \frac{1}{2} r_H g^2_w w_\mu w^\mu K_H
\eea
where $m^2_w=6 r_\phi g^2_w M^2_P$. For $r_\phi=r_H=1$, we recover the simple Lagrangian introduced  in eq.~(\ref{weylinv}) in the text.

In this case, making a field redefinition of the Weyl gauge field by
\bea
{\tilde w}_\mu &=&w_\mu-\frac{g_wr_H}{2}\cdot  \frac{1}{m^2_w+ g^2_w r_H K_H}\, K_\mu \nonumber \\
&=& w_\mu-\frac{1}{2g_w}\partial_\mu \ln (m^2_w+ g^2_w r_H K_H), \label{wredef}
\eea
we obtain the general gauge-fixed Lagrangian as
\bea
\frac{{\cal L}}{\sqrt{-g}} &=& -\frac{1}{2}  (M^2_P+2\xi_H |H|^2) R +6\xi_H(r_H-1)|\partial_\mu H|^2+(1-6\xi_H(r_H-1))  |D'_\mu H|^2  \nonumber \\
&& -V(H)- \frac{1}{4} X_{\mu\nu} X^{\mu\nu} -\frac{1}{4} {\tilde w}_{\mu\nu} {\tilde w}^{\mu\nu}  \nonumber \\
&&+ \frac{1}{2}(m^2_w + g^2_w r_H K_H){\tilde w}_\mu {\tilde w}^\mu -  \frac{ g^2_w r^2_H}{8} \,\frac{K_\mu K^\mu}{m^2_w + g^2_w r_H K_H}.  \label{gf2g}
\eea
Here, we find that  $m^2_w + g^2_w r_H K_H=6r_\phi g^2_w M^2_P(1+2(r_H/r_\phi) \xi_H |H|^2/M^2_P)$.
As a result, we also get the general Einstein-frame Lagrangian as
\bea
\frac{{\cal L}_E}{\sqrt{-g_E}} &=&-\frac{M^2_P}{2} R+6\xi_H(r_H-1) \frac{|\partial_\mu H|^2}{\Omega}+ (1-6\xi_H(r_H-1)) \frac{|D'_\mu H|^2}{\Omega}  \nonumber \\
&&+\frac{3\xi^2_H}{M^2_P} \frac{(\partial_\mu |H|^2)^2}{\Omega^2}- \frac{V(H)}{\Omega^2}- \frac{1}{4} X_{\mu\nu} X^{\mu\nu} -\frac{1}{4} {\tilde w}_{\mu\nu} {\tilde w}^{\mu\nu} \nonumber \\
&&+ \frac{1}{2}(m^2_w + g^2_w r_H K_H)  \Omega^{-1} {\tilde w}_\mu {\tilde w}^\mu-  \frac{ g^2_w r^2_H}{8\Omega} \,\frac{K_\mu K^\mu}{m^2_w + g^2_w r_H K_H}.
\eea
So, for the general Weyl-invariant Lagrangian, the non-canonical Higgs kinetic terms are not cancelled completely, and the Weyl gauge field $ {\tilde w}_\mu$ is not decoupled from the Higgs field.

For simplicity, we take the case with $r_H=r_\phi$. Then, from $m^2_w + g^2_w r_H K_H=6r_\phi g^2_w M^2_P \Omega$, a simplification arises, as follows,
\bea
\frac{{\cal L}_E}{\sqrt{-g_E}} &=&-\frac{M^2_P}{2} R+6\xi_H(r_H-1) \frac{|\partial_\mu H|^2}{\Omega} + (1-6\xi_H(r_H-1)) \frac{|D'_\mu H|^2}{\Omega}\nonumber \\
&+&\frac{3\xi^2_H(1-r_H)}{M^2_P} \frac{(\partial_\mu |H|^2)^2}{\Omega^2}  - \frac{V(H)}{\Omega^2}- \frac{1}{4} X_{\mu\nu} X^{\mu\nu} -\frac{1}{4} {\tilde w}_{\mu\nu} {\tilde w}^{\mu\nu} + \frac{1}{2} m^2_w {\tilde w}_\mu {\tilde w}^\mu. \label{genweyl}
\eea
In this case, we find that the non-canonical Higgs kinetic terms are not cancelled out, but the Weyl gauge field $ {\tilde w}_\mu$ is decoupled from the Higgs field.

\end{document}